# Spatiotemporal modulation of surface texture for information encoding and object manipulation

Xiao Yang[1], Jay Sim[1] & Ruike Renee Zhao[1]*

[1]Department of Mechanical Engineering, Stanford University, Stanford, CA 94305, USA.
*Corresponding author. Email: rrzhao@stanford.edu

**Abstract:** Dynamically tunable surface textures offer a powerful route to spatiotemporally regulate surface and interfacial properties, enabling emerging applications ranging from adaptive optics to soft robotic manipulation. However, achieving programmable, reversible, and spatiotemporal modulation of surface texture remains a fundamental challenge. Here, we present a photothermal-actuated liquid crystal elastomer bilayer that enables reversible, on-demand spatiotemporal modulation of surface textures through dynamically emerging and propagating wrinkles. Using direct laser writing or projected light fields, programmable and self-erasable wrinkle patterns are generated for dynamic information encoding. This spatiotemporal wrinkling enables object manipulation across diverse geometries, including uphill transport and navigation along predesigned paths. By coupling wrinkle-driven motion with thermally reversible dynamic bonding, the bilayer further enables assembly and disassembly of dynamic polymers, as well as cargo transportation. This work demonstrates spatiotemporally programmable wrinkling as a powerful mechanism for dynamic modulation of surface textures, establishing a versatile platform for multifunctional and reconfigurable smart surfaces.

# Introduction

Dynamically tunable surface textures offer a powerful route to actively regulate surface and interfacial properties, including optical reflectance, friction, adhesion, and wettability, in both space and time[1-3]. Such spatiotemporal control is increasingly essential for emerging technologies ranging from adaptive optics and reconfigurable information displays to soft robotic locomotion and object manipulation[4-8], where surface functionality must evolve locally, reversibly, and on demand. Despite its importance, achieving programmable spatiotemporal modulation of surface texture remains a fundamental challenge, as most existing approaches suffer from limited reversibility, coarse spatial resolution, or static patterning.

Surface wrinkling offers a uniquely attractive platform for dynamic surface texture programming because it converts simple mechanical instabilities into rich, reversible, and tunable surface morphologies[9]. Widely observed in natural systems such as human skin, fruit, and biological cortices, wrinkling supports critical functions including surface hydration regulation, mechanical protection, enhanced grip, and increased interfacial area for biochemical activity[10-12]. In engineered materials, wrinkling arises when compressive strain is applied to a stiff thin film bonded to a compliant substrate, causing the film to buckle out of plane once a critical strain threshold is exceeded[13-15]. When the strain is applied reversibly, wrinkles can emerge, evolve, and disappear, enabling real-time modulation of surface properties for applications in smart windows[16], droplet manipulation[17,18], adaptive camouflage[19,20], and information encoding[21,22].

Achieving spatiotemporal wrinkling beyond globally actuated static patterns enables advanced surface functions that require not only static or switchable textures, but also surface patterns that can be written, erased, and transported in a controlled manner. This dynamic modulation of surface textures provides a versatile platform for directional transport, programmable object interaction, and adaptive surface responses, where functionality arises from the controlled motion of surface features rather than from surface morphology alone. Realizing spatiotemporal wrinkling requires a mechanism capable of generating localized and reversible compressive strain with high spatial and temporal resolution. Among various external stimuli, light is especially attractive because it enables rapid, wireless, and non-contact actuation while offering exceptional spatiotemporal programmability. Illumination can be patterned, scanned, or dynamically modulated to selectively activate specific regions of a surface, allowing wrinkles to be generated only where and when needed.

Here, we present spatiotemporally programmable wrinkling as a powerful strategy for surface pattern regulation, information encoding, and object manipulation through dynamically emerging and propagating wrinkles. We engineer a photothermal-responsive liquid crystal elastomer (LCE) bilayer, actuated through direct light writing or projected illumination, to enable on-demand, reversible, and spatiotemporal surface wrinkling. As shown in **Fig. 1a**, the LCE bilayer is fabricated by spin-coating LCE ink onto a pre-stretched LCE substrate, followed by UV curing to form a stiff LCE film on top of a soft LCE substrate. Upon localized light irradiation, the LCE substrate undergoes photothermally driven nematic-isotropic phase transition[23,24], inducing localized contraction and wrinkle formation in the irradiated region. Spatially programmable wrinkle patterns can then be generated by laser writing or projected light for self-erasable information encoding (**Fig. 1b**). Beyond patterning, translating the light stimulus produces dynamically propagating wrinkles that drive controlled object manipulation (**Fig. 1c**). These capabilities further enable higher-level functionalities, including reversible assembly and disassembly of dynamic polymers (**Fig. 1d**,i), and wrinkle-driven cargo transportation (**Fig. 1d**,ii). Together, this work demonstrates spatiotemporally programmable wrinkling as a powerful mechanism for dynamic and reversible modulation of surface textures, establishing a versatile platform for multifunctional and reconfigurable smart surfaces.

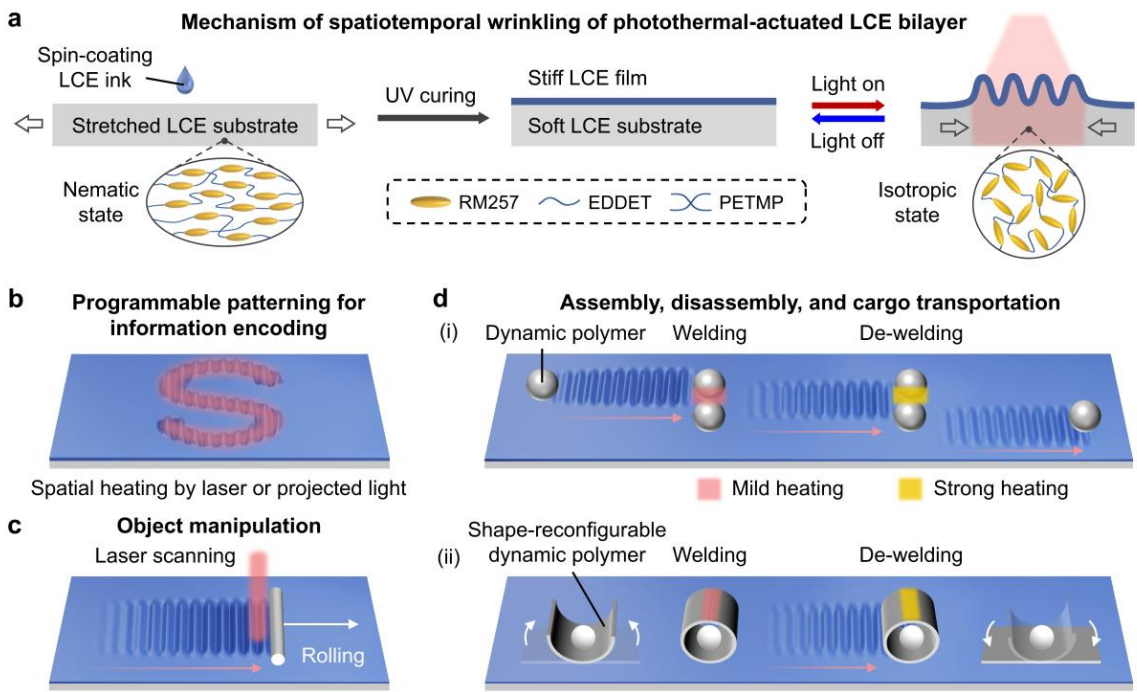

**Fig. 1. Mechanism and applications of spatiotemporal wrinkling of photothermal-actuated LCE bilayer. a**, Schematics of the fabrication process and working mechanism of the LCE bilayer. Localized photothermal heating induces contraction of the LCE substrate through the nematic-isotropic phase transition, resulting in localized wrinkle formation on the LCE film. **b-d**, Applications of spatiotemporal wrinkling for programmable patterning for information encoding (**b**), object manipulation (**c**), and dynamic polymer-enabled assembly, disassembly, and cargo transportation (**d**).

## Results and discussion

### Thermomechanical characterization of spatiotemporal wrinkling of LCE bilayers

The wrinkle wavelength, amplitude, and orientation are governed by well-defined mechanical parameters, including film and substrate moduli, film thickness, and applied strain[25]. **Figure 2a** defines the key geometric and mechanical parameters of the LCE bilayer. Importantly, both left and right ends of the bilayer are clamped to prevent global contraction of the bilayer upon heating. In this work, all presented bilayers share the same mechanical properties with the film Young's modulus being $E_f = 7.24$ MPa and substrate Young's modulus being $E_s = 1.12$ MPa (Supplementary **Fig. 1**). However, the film thickness $H_f$ varies for each bilayer to tune the wrinkle wavelength and amplitude. The peak-to-peak wavelength $\lambda$ and peak-to-valley amplitude $A$, which evolve with thermally induced substrate contraction, are used to characterize the wrinkle morphology. The chemical formulations of the LCE bilayer are provided in Supplementary **Fig. 2**, and additional details on the fabrication process are available in the **Methods**. Wrinkling begins when the local temperature exceeds the critical wrinkling temperature of $T_w = 55°C$, at which the thermally induced contraction of the LCE substrate accumulates sufficient compressive strain to trigger mechanical instability[26] (Supplementary **Fig. 3**). **Figure 2b** shows the temperature distribution on the illuminated region of the film. An IR laser is applied to the bilayer surface with the lit area marked by dashed white lines (**Fig. 2b**,i). The corresponding measured temperature is plotted in **Fig. 2b**,ii, with the surface temperature increasing toward the center of the lit area and reaching a maximum of 82.6°C. The wrinkling area slightly extends beyond the illuminated area due to heat transfer, while remaining spatially confined around the light-exposed region.

Film thickness is modulated to control the wrinkle resolution, as seen in **Fig. 2c**. Here, the film thickness is controlled during fabrication by adjusting the spin-coating speed of the LCE ink, producing LCE bilayers with $H_f = 0.08$ mm, 0.11 mm, and 0.14 mm. Increasing film thickness leads to larger wrinkle wavelengths and thus reduces the spatial resolution. The finite element analysis (FEA) results indicate the maximum in-plane principal strain distribution, illustrating strong agreement between simulation results and experimental images of wrinkling morphology (the FEA implementation is shown in Supplementary **Note 1**). The temporal evolution of localized wrinkles during photothermal irradiation and the corresponding temperature dependence of wrinkle wavelength and amplitude are shown in Supplementary **Fig. 4** and **Fig. 5**. **Figure 2d** quantifies the effect of film thickness on wrinkle wavelength and amplitude at the fully developed wrinkled state (maximum LCE substrate contraction in the heated region). Both wavelength and amplitude increase linearly with increasing film thickness, in good agreement with FEA predictions. These results demonstrate that tuning film thickness provides an effective means to tailor surface wrinkling of LCE bilayers, enabling programmable control over wrinkle morphology and establishing a versatile design framework for customizable surface textures.

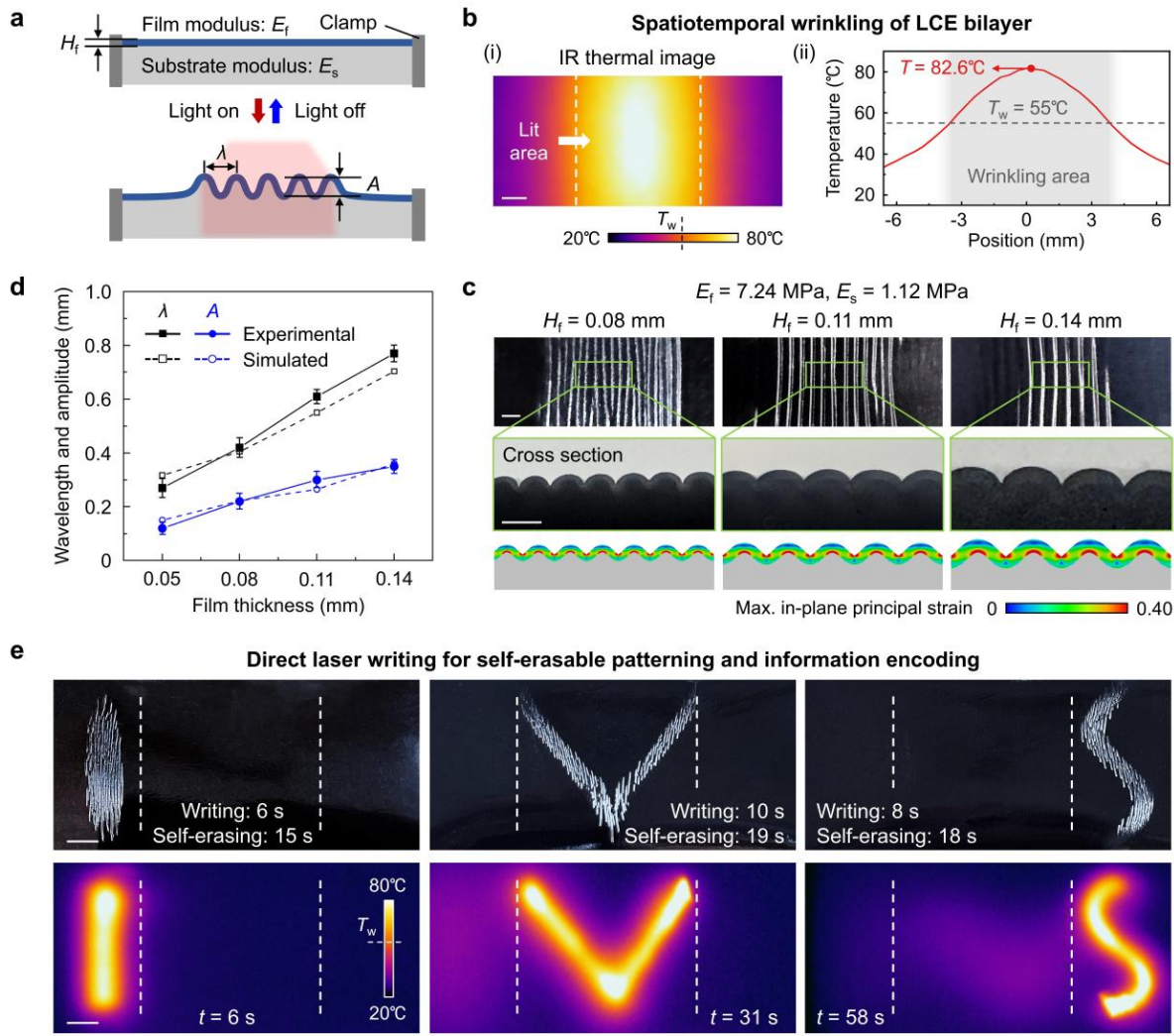

**Fig. 2. Thermomechanical characterization of spatiotemporal wrinkling of LCE bilayers and laser writing for self-erasable patterning and information encoding. a**, Schematics of spatiotemporal wrinkling of the LCE bilayer, with definitions for film thickness $H_f$, Young's moduli of the film and substrate $E_f$ and $E_s$, wavelength $\lambda$, and amplitude $A$. **b**, IR thermal image and temperature profile of LCE film surfaces showing the spatial temperature distribution under irradiation. Scale bar: 1 mm. **c**, Experimental images of LCE film surfaces showing the wrinkle morphologies under irradiation, together with corresponding experimental and FEA cross-sectional profiles. All bilayers share the same $E_f$ = 7.24 MPa and $E_s$ = 1.12 MPa but different $H_f$ = 0.08 mm, 0.11 mm, and 0.14 mm. Scale bars: 1 mm for film surfaces and 0.5 mm for cross sections. **d**, Wavelength (black) and amplitude (blue) of LCE bilayers as a function of film thickness. Data are shown as mean ± SD (n = 3). **e**, Optical and IR thermal images of I-, V-, and S-shaped wrinkle patterns generated by direct laser writing. Patterns are written sequentially from left to right on the bilayer surface, with white dashed lines indicating boundaries between adjacent writing regions. The wrinkle patterns spontaneously self-erase upon cooling, enabling transient and reconfigurable information encoding. Scale bars: 3 mm.

**Laser writing for self-erasable patterning and information encoding**

Wrinkling-based surface patterning provides a powerful strategy for designing intelligent materials with dynamic display capabilities[27]. **Figure 2e** shows direct laser writing of self-erasable wrinkle patterns for dynamic information encoding. An IR laser is scanned along a programmed trajectory, enabling real-time generation of localized wrinkles that follow the laser path. Sequential laser irradiation produces "I", "V", and "S" patterns from left to right on the bilayer, all of which appear within 10 s. These patterns disappear spontaneously within 20 s as the localized heat dissipates, fully reversing the wrinkle formation (Supplementary **Video 1**). These results demonstrate that direct laser writing enables rapid, on-demand formation of diverse patterns. The progressive emergence of wrinkles along the laser-scanning path highlights the potential for real-time and reconfigurable optical display applications. When the wrinkle patterns need to be kept in place, patterned illumination by a projector can maintain the patterns for a desired duration. See Supplementary **Fig. 6** for programmable wrinkle patterning enabled by projected light and Supplementary **Fig. 7** for the effect of film thickness on the projected wrinkle patterns.

**Dynamically propagating wrinkles for object manipulation**

Beyond programmable patterning, spatiotemporal thermal actuation of the LCE bilayer can also generate propagating wrinkles for object manipulation. **Figure 3a**,i shows that a ball (mass: 20 mg, used in all subsequent demonstrations) placed on an LCE bilayer ($H_f$ = 0.14 mm) can be propelled forward by laser-induced, dynamically propagating wrinkles. The laser-illuminated area is denoted by the red box, which continuously follows the ball as it moves across the bilayer surface. Here, localized photothermal heating generates a wrinkled region behind the ball, raising the film and thereby causing the ball to roll forward and across the bilayer surface. The ball travels 20 mm within 21 s, resulting in an average rolling speed of 0.95 mm·s$^{-1}$. In addition, the propagating wrinkles can also propel objects with diverse geometries. As shown in **Fig. 3a**,ii, a rod (mass: 80 mg) is driven to roll forward on the bilayer, with a rolling speed of 0.68 mm·s$^{-1}$. In contrast, a bolt (mass: 650 mg) can be rotated when the region adjacent to the bolt head is laser-heated. The localized wrinkles interact primarily with the head region, while the tip acts as a pivot, inducing a rotation of 120° within 56 s (**Fig. 3a**,iii). Note that for the rotation of a bolt, a thinner bilayer ($H_f$ = 0.05 mm) is used to generate finer wrinkles that offer better localized actuation control required for rotation. Actuation of all three objects can be seen in Supplementary **Video 2**.

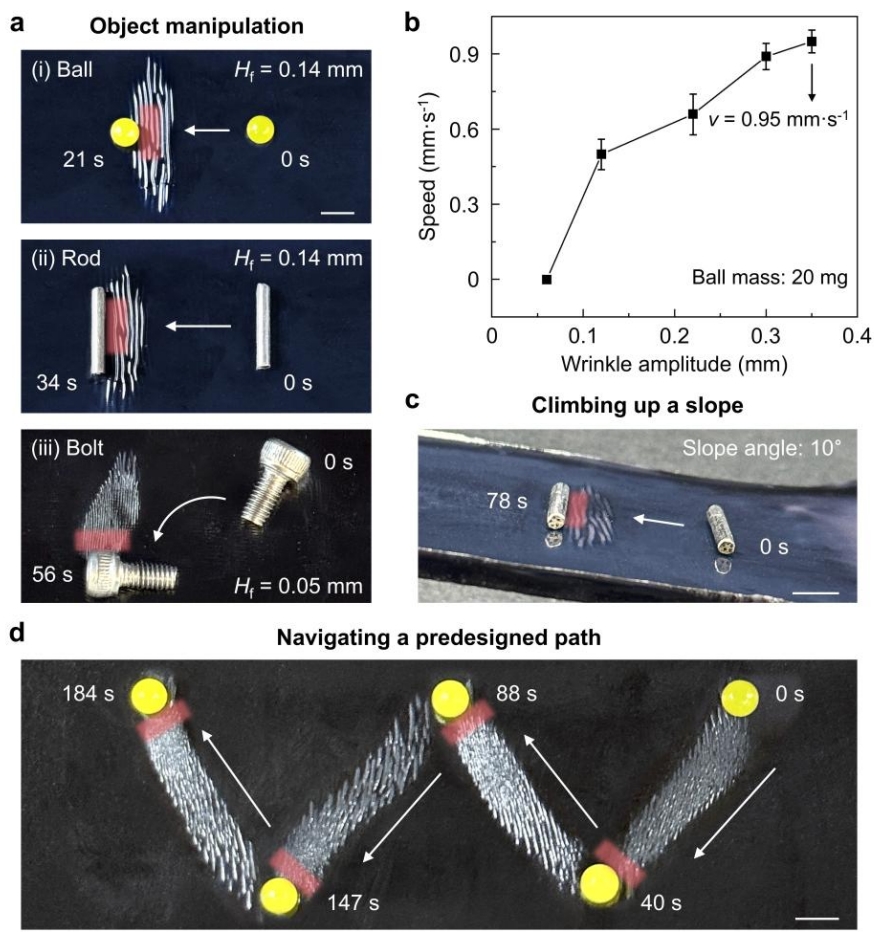

**Fig. 3. Dynamically propagating wrinkles for object manipulation. a**, Experimental images illustrating forward rolling of (i) a ball and (ii) a rod, and (iii) rotary rolling of a bolt enabled by dynamically propagating wrinkles. The IR laser continuously follows behind the objects and moves along with them, generating dynamically propagating wrinkles that drive objects' rolling motion. **b**, The rolling speed of the ball as a function of wrinkle amplitude on LCE bilayers. Data are shown as mean ± SD (n = 3). **c**, A rod climbing up a slope with a slope angle of 10°. **d**, A ball navigating a predesigned path. Scale bars: 5 mm.

The rolling speed of the object depends on the wrinkle amplitude of the bilayer, which is governed by the film thickness (**Fig. 2d**). **Figure 3b** plots the rolling speed of the ball on bilayers versus wrinkle amplitude. For amplitudes below $A = 0.06$ mm, the generated wrinkles possess an insufficient surface height, and therefore no rolling is observed. For $A = 0.12$ mm, 0.22 mm, 0.30 mm, and 0.35 mm, the corresponding average rolling speeds are 0.50 mm·s$^{-1}$, 0.66 mm·s$^{-1}$, 0.89 mm·s$^{-1}$, and 0.95 mm·s$^{-1}$, respectively. As the amplitude increases, the ball rolls faster due to the steeper surface slopes from wrinkles with larger amplitudes that generates larger driving forces. However, achieving larger amplitudes to further increase the rolling speeds is challenging, because increasing the film thickness beyond a certain point promotes global bending actuation rather than localized wrinkling. Larger amplitudes also enable manipulation of heavier objects (**Supplementary Table 1**). Note that increasing the object mass reduces rolling speed due to larger force being needed to propel the object (Supplementary **Fig. 8**). In addition to planar transport, the LCE bilayer ($H_f$ = 0.14 mm) can also drive a rod to climb an inclined surface with a slope angle up to 10° without slipping (**Fig. 3c**). **Figure 3d** demonstrates active, multi-directional control of object motion, in which dynamically propagating wrinkles steer a rolling ball along different directions by modulating the wrinkle path with a scanning laser. The ball navigates a predesigned W-shaped path on the bilayer ($H_f$ = 0.05 mm), closely following the scanning laser. Both demonstrations are shown in Supplementary **Video 2**. These results establish dynamically propagating wrinkling as a powerful platform for laser-driven object manipulation, enabling programmable transport on textured surfaces.

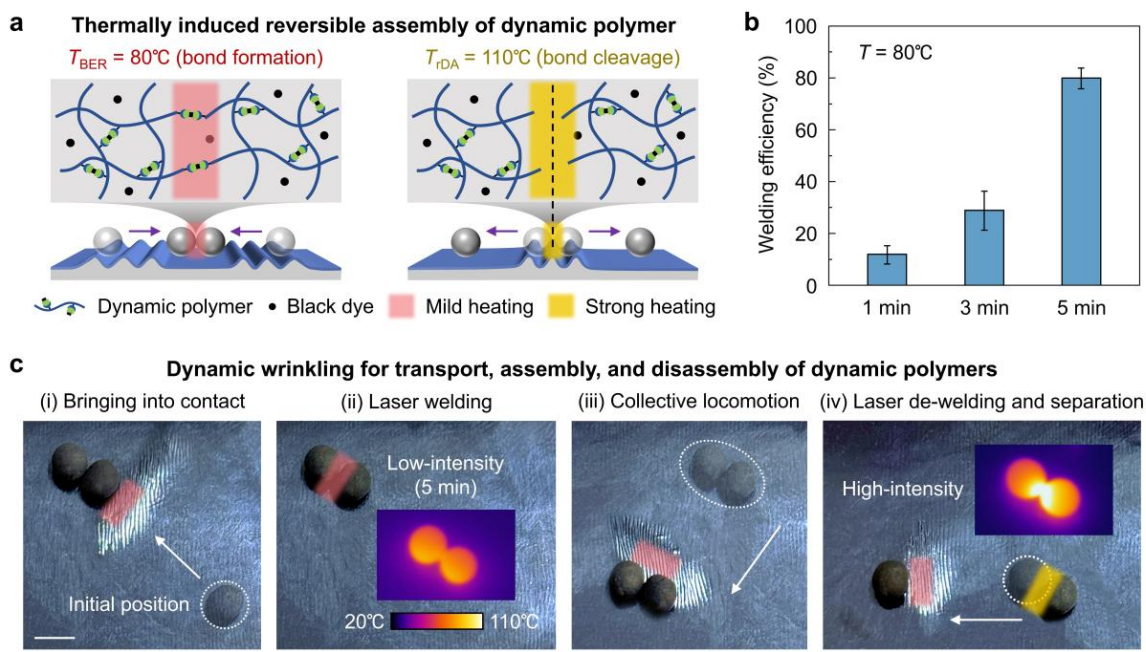

**Fig. 4. Dynamic wrinkling via propagation and emergence for transport, assembly, and disassembly of dynamic polymers. a**, Schematics of the working mechanism of thermally induced assembly and disassembly of dynamic polymers. For assembly, dynamically propagating wrinkles bring the polymer balls into contact, and low-intensity irradiation activates DA bond formation at the mild temperature $T_{BER}$ for welding. For disassembly, high-intensity irradiation triggers DA bond cleavage at the elevated temperature $T_{rDA}$, after which the emerging wrinkles at the interface between the polymer balls separate them. **b**, Welding efficiency of dynamic polymer balls after different heating times at 80°C. **c**, Experimental images illustrating assembly and disassembly process of dynamic polymers. Scale bar: 3 mm.

**Dynamic wrinkling via propagation and emergence for transport, assembly, and disassembly of dynamic polymers**

In the previous section, we demonstrated object locomotion driven by dynamically propagating wrinkles. Building on this capability, we extend the approach to enable reversible assembly and disassembly of dynamic polymers. **Figure 4a** illustrates the underlying mechanisms. The dynamic polymer used in this study is a thermally reversible Diels-Alder (DA) network based on furan-maleimide chemistry[28,29] (Supplementary **Fig. 9**). A black dye is incorporated into the polymer matrix to enable efficient photothermal heating.

To assemble, propagating wrinkle patterns first move the individual dynamic polymers such that they physically contact. Subsequent low-intensity laser irradiation induces mild heating at the interface of the dynamic polymers, activating the bond exchange reaction (BER) and welding the polymers together at $T_{BER}$ = 80°C. For disassembly, higher-intensity irradiation is applied at the polymer interface to trigger the retro Diels-Alder (rDA) reaction, leading to bond cleavage at the elevated temperature of $T_{rDA}$ = 110°C. Following bond breakage, the emergence of wrinkles at the polymer interface drives the separation. Both characteristic temperatures are adopted from the thermomechanical characterization of the same dynamic polymer system reported in our earlier work[29]. **Figure 4b** shows how the welding efficiency of dynamic polymers varies with heating time at 80°C. Here, the welding efficiency is defined as the ratio of the tensile fracture strain of the welded dynamic polymer to the pristine dynamic polymer, which is quantified from uniaxial tensile tests (Supplementary **Fig. 10**). After 5 min of heating at 80°C, the welding efficiency exceeds 80% and is therefore used as the welding duration in all subsequent demonstrations.

**Figure 4c** demonstrates assembly and disassembly of two dynamic polymer balls on an LCE bilayer ($H_f$ = 0.05 mm) (Supplementary **Video 3**). Initially, dynamically propagating wrinkles generated by an IR laser irradiation propel one polymer ball into contact with the other (**Fig. 4c**,i). The interface of the contacted pair (red region) is then irradiated at 80°C for 5 min, enabling stable interfacial bonding and forming a welded structure (**Fig. 4c**,ii). Importantly, the laser is applied laterally such that the film of the bilayer is not heated and wrinkles are not generated. Once welded, the assembled pair behaves as a single body and exhibits collective locomotion under continuous wrinkle-induced actuation without interfacial failure (**Fig. 4c**,iii). Disassembly is achieved by applying a higher intensity laser laterally, as shown by the yellow region, to raise the polymer interface temperature above 110°C and initiate DA bond cleavage. Immediately afterward, the laser is directed onto the bilayer surface between the two polymer balls to generate localized wrinkles, which push the two dynamic polymer balls apart, separating the pair into two individual components (**Fig. 4c**,iv).

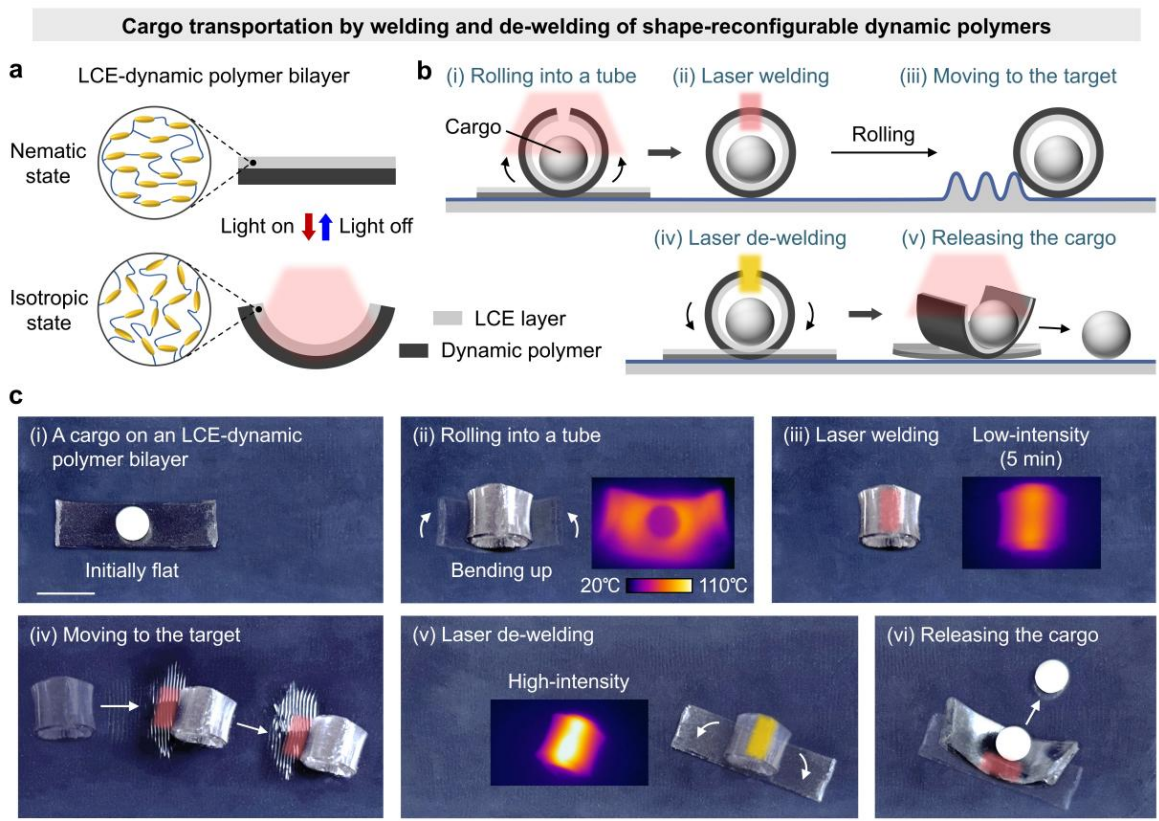

**Fig. 5. Cargo transportation by welding and de-welding of shape-reconfigurable dynamic polymers. a**, Schematics of the photothermally reversible bending of the LCE-dynamic polymer bilayer. **b,c**, Schematics and experimental images illustrating the wrinkle-driven cargo transportation process enabled by welding and de-welding of a shape-reconfigurable LCE-dynamic polymer bilayer. Scale bar: 10 mm.

**Cargo transportation by welding and de-welding of shape-reconfigurable dynamic polymers**

Building on the wrinkle-induced actuation and reversible bonding of dynamic polymers demonstrated above, we introduce a multifunctional LCE-dynamic polymer bilayer that integrates shape reconfiguration, welding and de-welding, and wrinkle-driven locomotion, enabling controlled cargo transportation under a single photothermal stimulus. As shown in **Fig. 5a**, a photothermal-responsive LCE-dynamic polymer bilayer is fabricated, consisting of a deformable LCE layer on top of a dynamic polymer layer that enables reversible welding, as detailed in the **Methods**. Upon irradiation, the LCE layer contracts through the nematic-isotropic phase transition, while the dynamic polymer layer remains inactive, causing the bilayer to bend toward the LCE layer and subsequently roll into a tubular structure around the target object (**Fig. 5b**,i). Laser welding of the outer dynamic polymer layer converts the bilayer into a closed carrier (**Fig. 5b**,ii) that can be transported by dynamically propagating wrinkles (**Fig. 5b**,iii). Subsequent laser de-welding removes the interfacial bonding constraint, allowing the stored elastic energy in the LCE layer to be released, which drives reopening of the carrier (**Fig. 5b**,iv). The following photothermally induced deformation of the bilayer releases the cargo (**Fig. 5b**,v), completing the transportation process.

**Figure 5c** presents the corresponding experimental process (Supplementary **Video 4**). A spherical cargo (mass: 80 mg) is placed on a flat LCE-dynamic polymer bilayer (25 mm × 8 mm × 1.5 mm) (**Fig. 5c**,i). Laser irradiation induces upward bending of the bilayer and formation of a tubular structure (**Fig. 5c**,ii). When the two edges meet, low-intensity irradiation at 80°C for 5 min activates DA bond formation at the contacting interface, welding the edges together (**Fig. 5c**,iii). The welded tube on the LCE bilayer ($H_f$ = 0.14 mm) is then propelled forward by propagating wrinkles, transporting the cargo toward the target (**Fig. 5c**,iv). Higher-intensity irradiation (> 110°C) triggers DA bond cleavage, leading to de-welding of the edges and reopening of the tube (**Fig. 5c**,v). Finally, asymmetric photothermal bending is achieved by selectively heating one side of the bilayer surface, which causes the bilayer to bend laterally and guides the cargo to roll out of the LCE-dynamic polymer bilayer (**Fig. 5c**,vi). These demonstrations highlight dynamically propagating wrinkling as a versatile actuation strategy that enables controlled locomotion, reversible assembly, and cargo transportation when coupled with dynamic polymers.

## Conclusion

In summary, we have developed a photothermal-actuated LCE bilayer that allows on-demand, reversible, and spatiotemporal modulation of surface textures through dynamically emerging and propagating wrinkles. Spatially programmable and self-erasable wrinkle patterns can be generated using direct laser writing or projected light fields, providing a versatile route for dynamic information encoding. This spatiotemporal wrinkling strategy enables object manipulation across a wide range of geometries and controlled locomotion including uphill transport and precise navigation along predesigned paths, highlighting the versatility of the LCE bilayer system. Moreover, the dynamically emerging and propagating wrinkles facilitate assembly and disassembly of dynamic polymers through their thermally reversible bonding interactions. This reversible assembling capability further enables cargo transportation using a shape-reconfigurable dynamic polymer, whose locomotion is driven by dynamically propagating wrinkles. Overall, this work establishes spatiotemporally programmable wrinkling as a powerful mechanism for dynamic and reversible surface texture modulation, opening new opportunities for the development of multifunctional and reconfigurable smart surfaces.

## Methods

**Fabrication of LCE substrate**

To fabricate the LCE substrate, the di-acrylate liquid crystal monomer 1,4-Bis-[4-(3-acryloyloxypropyloxy)benzoyloxy]-2-methylbenzene (RM257, Kindchem (Nanjing) Co., Ltd, China) was dissolved in toluene (50.0 wt%) at 80°C for 5 min. After the mixture was cooled to room temperature, the di-thiol chain extender 2,2-(ethylenedioxy) diethanethiol (EDDET, Sigma Aldrich, USA) (25.0 wt%), the tetra-thiol crosslinker pentaerythritol tetrakis(3-mercaptopropionate) (PETMP, Sigma Aldrich, USA) (4.5 wt%), the photoinitiator Irgacure 819 (Sigma Aldrich, USA) (2.0 wt%), and the photothermal black dye (nigrosine, Sigma Aldrich, USA) (0.1 wt%) were introduced to the mixture. The mixture was then stirred for 3 min with a magnetic stir bar for homogenization, and the catalyst dipropylamine (DPA, Sigma Aldrich, USA) (0.8 wt%) was added to the mixture. After stirring and degassing, the mixture was poured into a polydimethylsiloxane (PDMS) mold to undergo Michael addition reaction at room temperature for 12 h and was heated at 80°C for 24 h to evaporate the solvent. After demolding, the lightly crosslinked LCE substrate was obtained. All mass ratios are given relative to RM257.

**Fabrication of LCE ink for the LCE film**

To fabricate the LCE ink, RM257 was dissolved in toluene (80.0 wt%) at 80°C for 5 min. After the mixture cooled to room temperature, EDDET (24.0 wt%) was introduced to the mixture and stirred for 3 min for homogenization. Then, Irgacure 819 (2.0 wt%) and DPA (0.4 wt%) were added to the mixture. After stirring for 1 min, the LCE ink was obtained. All mass ratios are given relative to RM257.

**Fabrication of LCE bilayer**

To fabricate the LCE bilayer, the LCE ink was spin-coated onto a uniaxially pre-stretched LCE substrate, which was exposed to air for 1 h at room temperature. The bilayer was obtained after being irradiated by UV light (385 nm) for 5 min to cure both the LCE film and the LCE substrate and bond them together.

**Synthesis of dynamic polymer**

The dynamic polymer was fabricated following our previous study[29]. First, furan-grafted prepolymer was synthesized. The epoxy oligomer poly(ethylene glycol) diglycidyl ether (PEGDE, average molecular mass: 500 g·mol$^{-1}$, Sigma Aldrich, USA), the chain extender furfurylamine (Sigma Aldrich, USA) (19.4 wt%), and the antioxidant 2,6-di-tert-butyl-4-methylphenol (Sigma Aldrich, USA) (0.8 wt%) were mixed in dimethylformamide (30.0 wt%). After degassing, the mixture was sealed and stirred at 80°C for 20 h, followed by heating at 110°C for another 4 h. The mixture was then cooled to room temperature, yielding a viscous light-yellow furan-grafted prepolymer solution. All mass ratios are given relative to PEGDE.

Then, dynamic polymer was synthesized by crosslinking the furan-grafted prepolymer with bismaleimide crosslinker. Briefly, the prepolymer solution, the crosslinker bismaleimide (Sigma Aldrich, USA) (10.0 wt%), and the black dye (0.05 wt%) were manually mixed to obtain a homogeneous precursor. The precursor was poured into a PDMS mold to undergo DA reaction at 50°C for 2 h and was heated at 70°C for 24 h to evaporate the solvent. After demolding, the dynamic polymer was obtained. All mass ratios are given relative to the prepolymer solution. To obtain dynamic polymer with a spherical shape, the sample was manually reshaped on a hot plate at 140°C.

### Fabrication of LCE-dynamic polymer bilayer

To fabricate the LCE-dynamic polymer bilayer, a lightly crosslinked LCE layer was first prepared using the same procedure described above for the LCE substrate. The LCE layer was then uniaxially pre-stretched and cured by UV light (385 nm) for 5 min. After curing, the two ends of the LCE layer were clamped, and a PDMS mold was placed on top. The dynamic polymer precursor was poured into the PDMS mold to undergo DA reaction at 50°C for 2 h and was heated at 70°C for 24 h to evaporate the solvent. After demolding, the LCE-dynamic polymer bilayer was obtained.

### Material characterizations and temperature measurements

Stress-strain curves of the LCEs and dynamic polymers, and the contraction strain versus temperature relationship of the LCE substrate are conducted using a dynamic mechanical analyzer (DMA 850, TA instruments, USA). For the tensile tests, LCE samples (30 mm × 10 mm × 1 mm) are stretched at a rate of 0.01 s$^{-1}$ along their longitudinal direction at 25°C. The resulting test datasets are fitted to a neo-Hookean model to obtain the shear moduli, $\mu$. Using the relationship, $E = 2\mu (1 + v)$, and assuming that the LCEs are incompressible with Poisson's ratio $v = 0.5$, the Young's moduli $E$ of the LCE film and LCE substrate are found to be 7.24 MPa and 1.12 MPa, respectively.

Temperature distributions of the LCE bilayers and dynamic polymers are recorded using an IR thermal camera (Fotric 348A, Fotric Inc., USA). The measurements are taken at 20 cm from the sample, and the camera thermal emissivity is calibrated as 0.95 by comparing the temperature measured with the thermal camera and a thermocouple placed on the sample surface. Ambient temperature and humidity are set as 25°C and 50%, respectively.

### Data availability

The data that support the findings of this study are available from the corresponding author upon request.

## Acknowledgments

We acknowledge financial support from the Army Research Office (ARO) ECP Award W911NF-23-1-0176 and Defense Advanced Research Projects Agency (DARPA) YFA D25AC00387.

## Author contributions

R.R.Z. conceived the research and supervised the study. X.Y. performed the experiments and data analysis. J.S. performed the FEA simulations. All authors wrote and edited the original draft.

## Competing interests

The authors declare no competing interests.